\documentclass{article}

\usepackage{arxiv}

\usepackage[utf8]{inputenc} 
\usepackage[T1]{fontenc}    
\usepackage{hyperref}       
\usepackage{url}            
\usepackage{booktabs}       
\usepackage{amsfonts}       
\usepackage{nicefrac}       
\usepackage{microtype}      
\usepackage{lipsum}		
\usepackage{graphicx}
\usepackage{natbib}
\usepackage{doi}
\usepackage{bm}
\usepackage[varvw]{newtxmath}       
\usepackage{amsmath}
\mathchardef\mhyphen="2D 
\usepackage{booktabs}
\usepackage{multirow}
\usepackage{comment}

\title{Data-driven multi-scale modeling and robust optimization of composite structure with uncertainty quantification}

\author{ 	{Kazuma ~Kobayashi} \\
	Nuclear Engineering and Radiation Science\\
	Missouri University of Science and Technology\\
	Rolla, MO 65409, USA \\
	\And
         {Shoaib ~Usman } \\
	Nuclear Engineering and Radiation Science\\
	Missouri University of Science and Technology \\
\And
         {Carlos ~Castano} \\
	Nuclear Engineering and Radiation Science\\
	Missouri University of Science and Technology \\
 \And
         {Dinesh ~Kumar } \\
	Department of Mechanical Engineering\\
	University of Bristol\\
	Bristol BS8 1TR, UK \\
 \And
      {Syed ~Alam} \\
	Nuclear Engineering and Radiation Science\\
	Missouri University of Science and Technology\\
	Rolla, MO 65409, USA \\
 }

\renewcommand{\shorttitle}{\textit{Handbook of Smart Energy Systems} }

\hypersetup{
pdftitle={A template for the arxiv style},
pdfsubject={q-bio.NC, q-bio.QM},
pdfauthor={David S.~Hippocampus, Elias D.~Striatum},
pdfkeywords={First keyword, Second keyword, More},
}

\begin{document}
\maketitle

\begin{abstract}
It is important to accurately model materials' properties at lower length scales (micro-level) while translating the effects to the components and/or system level (macro-level) can significantly reduce the amount of experimentation required to develop new technologies. Robustness analysis of fuel and structural performance for harsh environments (such as power uprated reactor systems or aerospace applications) using machine learning-based multi-scale modeling and robust optimization under uncertainties are required. The fiber and matrix material characteristics are potential sources of uncertainty at the microscale. The stacking sequence (angles of stacking and thickness of layers) of composite layers causes meso-scale uncertainties. It is also possible for macro-scale uncertainties to arise from system properties, like the load or the initial conditions. This chapter demonstrates advanced data-driven methods and outlines the specific capability that must be developed/added for the multi-scale modeling of advanced composite materials. This chapter proposes a multi-scale modeling method for composite structures based on a finite element method (FEM) simulation driven by
surrogate models/emulators based on microstructurally informed meso-scale materials models to study the impact of operational parameters/uncertainties using machine learning approaches. To ensure optimal composite materials, composite properties are optimized with respect to initial materials volume fraction using data-driven numerical algorithms. 
\end{abstract}


\section{Introduction}
Artificial intelligence and machine learning (AI/ML) have been identified as vital tools for the growth and sustainability of the nuclear industry \cite{khan2022digital,hassan2022machine,rahman2022leveraging}. These tools have the potential to accelerate the development and deployment of nuclear reactors, reduce costs, and provide faster solutions to a wider range of problems \cite{kobayashi2022practical,kobayashi2022uncertainty,kobayashi2022surrogate,verma2022reliability,kobayashi2022data,kobayashi2022DTAFQ}. In particular, the different U.S. agencies such as the Department of Energy (DOE), Department of Defense (DOD), and Air Force Office of Scientific Research (AFOSR) strongly recommend leveraging AI/ML techniques to solve problems such as structural design optimization, process optimization, and economic optimization, as well as manufacturing optimization, hazard detection, and non-nominal condition monitoring. This chapter will discuss structural optimization in consideration of various scales.

In the past few decades, multi-scale modeling has seen widespread use to improve engineering problems' accuracy, and efficiency \cite{kumar2020efficient}. As stated in our previous study \cite{kumar2020efficient}, ``in structural modeling, the performance of the composite can be affected by uncertainties in the material properties, model parameters, system loading, and manufacturing tolerances caused by the order in which the laminates are stacked. 
Statistical approaches can provide more confidence in computational results by incorporating their variability-bound and probabilistic behavior into the computation \cite{kumar2020efficient}." The goal of a simulation platform is to integrate different kinds of mathematical models, both commercially available and developed in the previous works \cite{kumar2020uncertainty, kumar2022multi,kumar2020nuclear,kumar2019influence,kumar2020efficient}. 

Uncertainties are inherent to the majority of material properties. Consequently, it is of the utmost importance to manage uncertainties and the propagation of their effects throughout the business decision workflow and material model. The complexity of nonlinear phenomena and the inherent unpredictability of materials modeling, materials processing, and business decision environments adds another layer of difficulty to the materials selection process (cost, business sustainability, etc.). In order to do this, statistical methods to measure how uncertain input parameters affect uncertain outputs are required  \cite{kumar2020uncertainty, kumar2022multi,kumar2020nuclear,kumar2019influence,kumar2020efficient}

In engineering design, the optimization of design parameters is an unavoidable task. Under various conditions, such as manufacturing cost and durability, their functions are optimized. The rapid evolution of computers has been a tremendous boon in engineering. Its application to engineering design is a prime example of its varied benefits. When design parameters are only partially known or are uncertain, they can be characterized by a probability distribution. 

The uncertainties in parameters are due to various factors such as manufacturing processes and material properties, and optimizing the final product's performance has traditionally required high financial cost and time. In order to address this issue, machine learning-based optimizations are applied in various industrial fields. Robust optimization is one of the methods for quantifying the robustness of a product's performance after taking its uncertain variables into account \cite{kumar2022multi,kumar2019influence}. 

In complex material design, the selection of materials is an important process based on various conditions such as performance, durability, and cost. Even when using a single substance, the microstructure of the substance will affect the final performance, which is one design variable. Design variables can be categorized by scale, defining micro, meso, and macro scales in descending order of decreasing size. The method of evaluating the impact of parameters at multiple scales on objective performance is called multi-scale modeling. Previously, the corresponding author has published a number of works on harsh  environments \cite{alam2019assembly,alam2019small1,alam2019small2,alam2019parametric,alam2019coupled,alam2020neutronic,alam2019neutronic,alam2020lattice,almutairi2022weight}, which require robust optimization in the multi-scale problem.

This chapter introduces robust optimization, an optimization technique that takes uncertainties into account. We have developed a simulation method and framework for composite materials that combine multi-scale modeling and robust optimization methods. 

It also discusses the treatment of problems where uncertainty variables affect multiple scales. Multi-scale modeling and robust carbon fiber composite design optimization are performed as a case study. As a test case, we modeled a carbon fiber laminated structure and performed robust optimization of the shear stress of the model.

\section{Robust optimization \& Multi-Scale Problem}
Traditional Optimization techniques tended to over-optimize solutions that perform well in theory but have poor off-design properties. In this manner, ensuring that the product's design requirements are met is crucial. The best option might not always be the most reliable one. Designers need to consider the system's ability to handle changes in operating conditions with minimal damage, alteration, or loss of functionality. This is known as the system's robustness \cite{modeFrontier}. 

\begin{figure}[!htbp]
    \centering
    \includegraphics[scale=0.7]{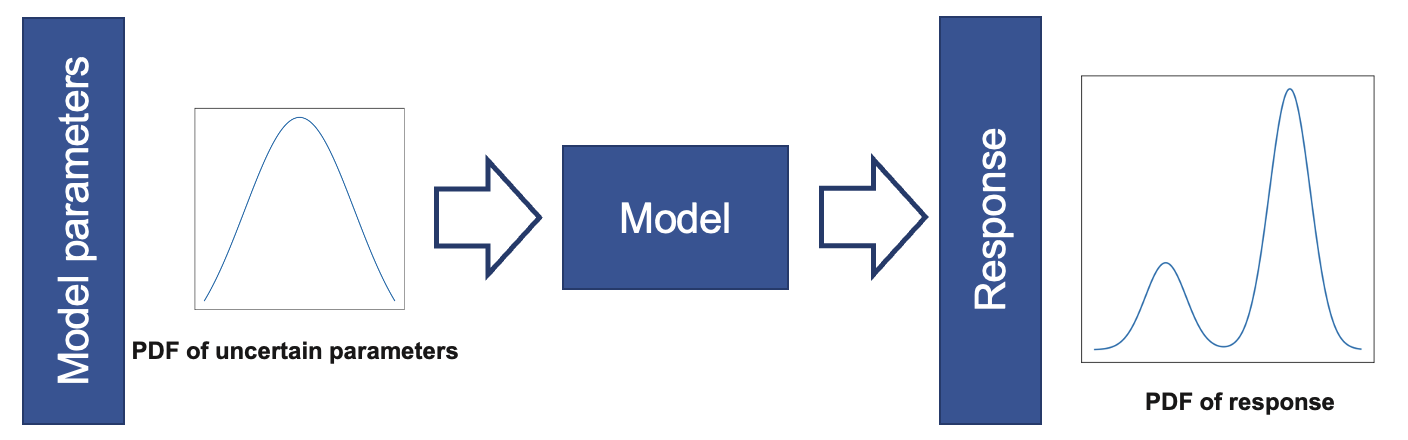}
    \caption{Simulation framework: stochastic inputs and system response}
    \label{fig:stochastic}
\end{figure}

Replacing deterministic models with stochastic models in optimization methods allows for handling uncertain (stochastic) input variables. The robust optimization handles the functions (or system responses) as a stochastic model. It fundamentally differs from the traditional deterministic model, which calculates a static response for a given input variable, as shown in Fig. \ref{fig:stochastic}. The statistical moments (mean and standard deviation) of that stochastic response are treated as the target value in robust optimization to optimize the problem. The statistical moments (mean and standard deviation) of that stochastic response are treated as the target value in robust optimization to optimize the problem.

One drawback of this approach is the computational cost. It is due to the need to perform calculations on an input variable's neighboring values to obtain statistical information such as the mean and standard deviation of the objective response. Thus, the statistical reliability of the optimal solution is strongly proportional to the computational cost.


Implementing robust optimization, including sensitivity for the multi-scale problems to observe the optimization and uncertainty over the length scales, which can significantly reduce the amount of experimentation required to develop new technologies. Complicated steps are involved in the creation of composite structures. Uncertainties of a composite structure's modeling process impact its performance at various scales. The microscale material properties of both the fiber and the matrix are responsible for the potential sources of uncertainty at this scale. Uncertainties in the mesoscale are caused by the order in which composite layers are stacked. Finally, on the macro-scale, the system properties, such as loading, boundary, and initial conditions, can lead to uncertainty. This chapter will implement the propagation of uncertainties from multiple scales using two composite applications. These uncertainties' effect on composite responses is provided using the non-intrusive approaches described in our previous works \cite{kumar2020uncertainty, kumar2022multi}.

\section{Case-Study: Multi-Scale Modeling with Uncertainty and Sensitivity}
A multi-scale modeling and robust optimization for a carbon fiber composite (AS4D/9310) are demonstrated. The problem setup is retrieved from Barbero and Ever J \cite{barbero2013finite}, "The model is a simply supported square plate $a_{x}=a_{y}=2000\,\rm mm$, $t=10\, \rm mm$ of thickness, laminated with AS4D/9319 in a $[0/90/\pm 45]_{S}$ symmetric laminated configuration. The plate is loaded with a tensile load $N_x = 100 \, \rm N/mm$ ($N_{y}=N_{xy}=M_{x}=M_{y}=M_{xy}=0$). The objective is to compute the in-plane shear stress in the lamina coordinate system $\sigma_{6}$ \cite{barbero2013finite}."

\begin{figure}[!htbp]
    \centering
    \includegraphics[scale=0.5]{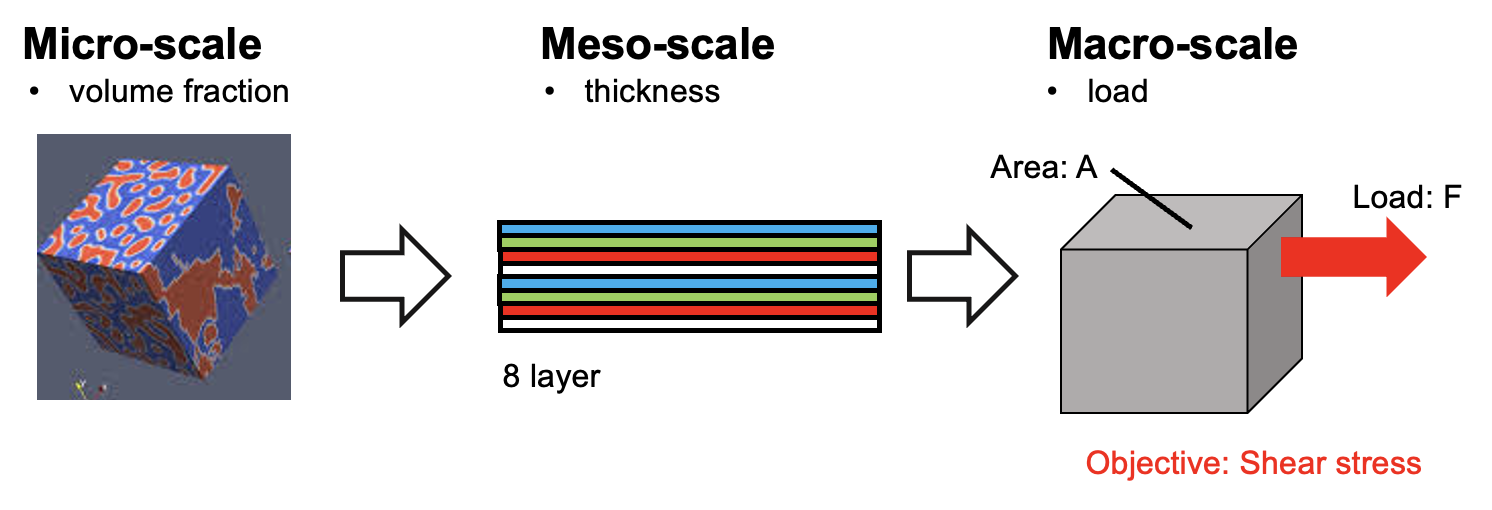}
    \caption{Multi-scale input variables for the carbon fiber composite model. In the meso-scale, white layers correspond to layer ID:1, red: 2, green: 3, and blue:4.}
    \label{fig:setup}
\end{figure}

In order to modify the problem into a multi-scale one, new input variables are set: a volume fraction of fiber phase (micro-scale), the thickness of layers (meso-scale), and load (macro-scale). Fig. \ref{fig:setup} shows a concept of this problem. Probability distributions for all input variables are assumed to follow a normal distribution with a standard deviation of 5\%. The lower and upper bounds of the range to search for input variables were also set, respectively. A summary of input variables is listed in Table \ref{tb:input}. 

The modeling and simulation are performed using the finite-element (FEM) solver ABAQUS. In the micro-scale simulation, the elastic properties, Young’s moduli, Poisson’s ratios, and shear modulus are computed with respect to a given volume fraction. The properties obtained from the micro-scale calculations are propagated to later simulations. In meso-scale, the thickness of each layer is treated as a variable, and they are stacked (to build a macro-scale model). Finally, the shear stress $\sigma_{6}$ is calculated when a tensile load $N_{x}$ is applied to the constructed model.

\begin{figure}[!htbp]
    \centering
    \includegraphics[scale=0.7]{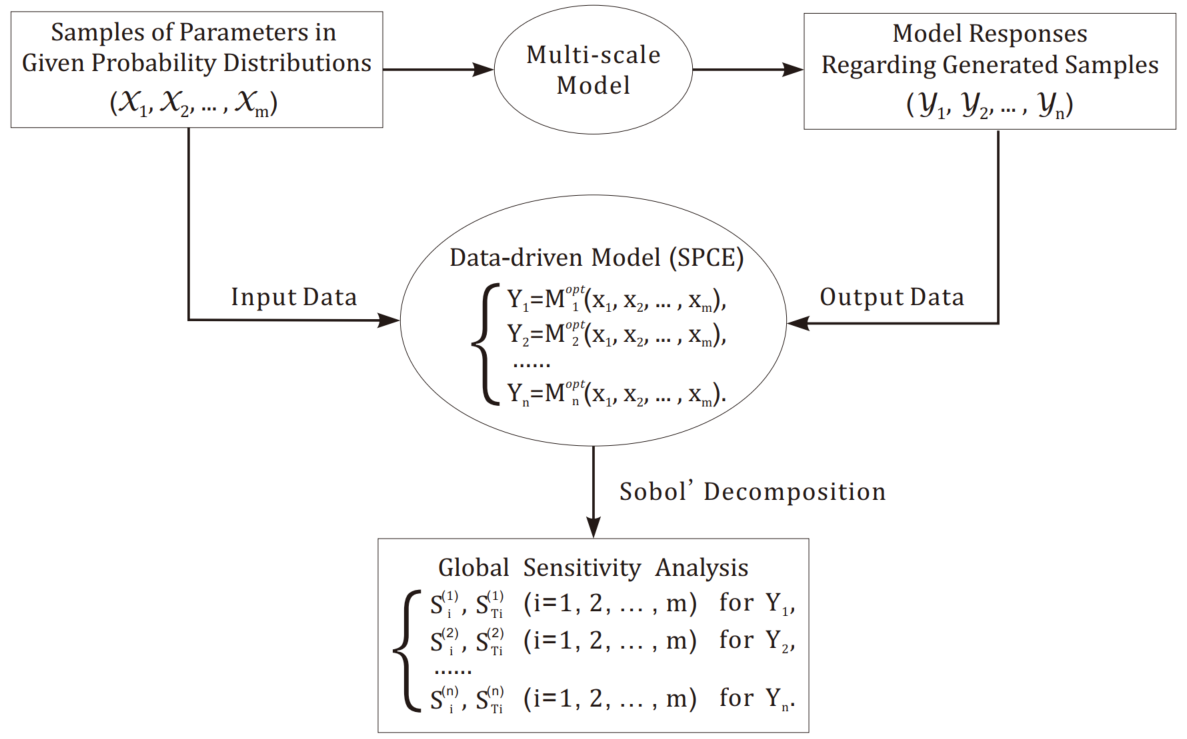}
    \caption{The process of data-driven analysis \cite{kumarLUXReport}}
    \label{fig:dd}
\end{figure}

\begin{figure}[!htbp]
    \centering
    \includegraphics[scale=0.30]{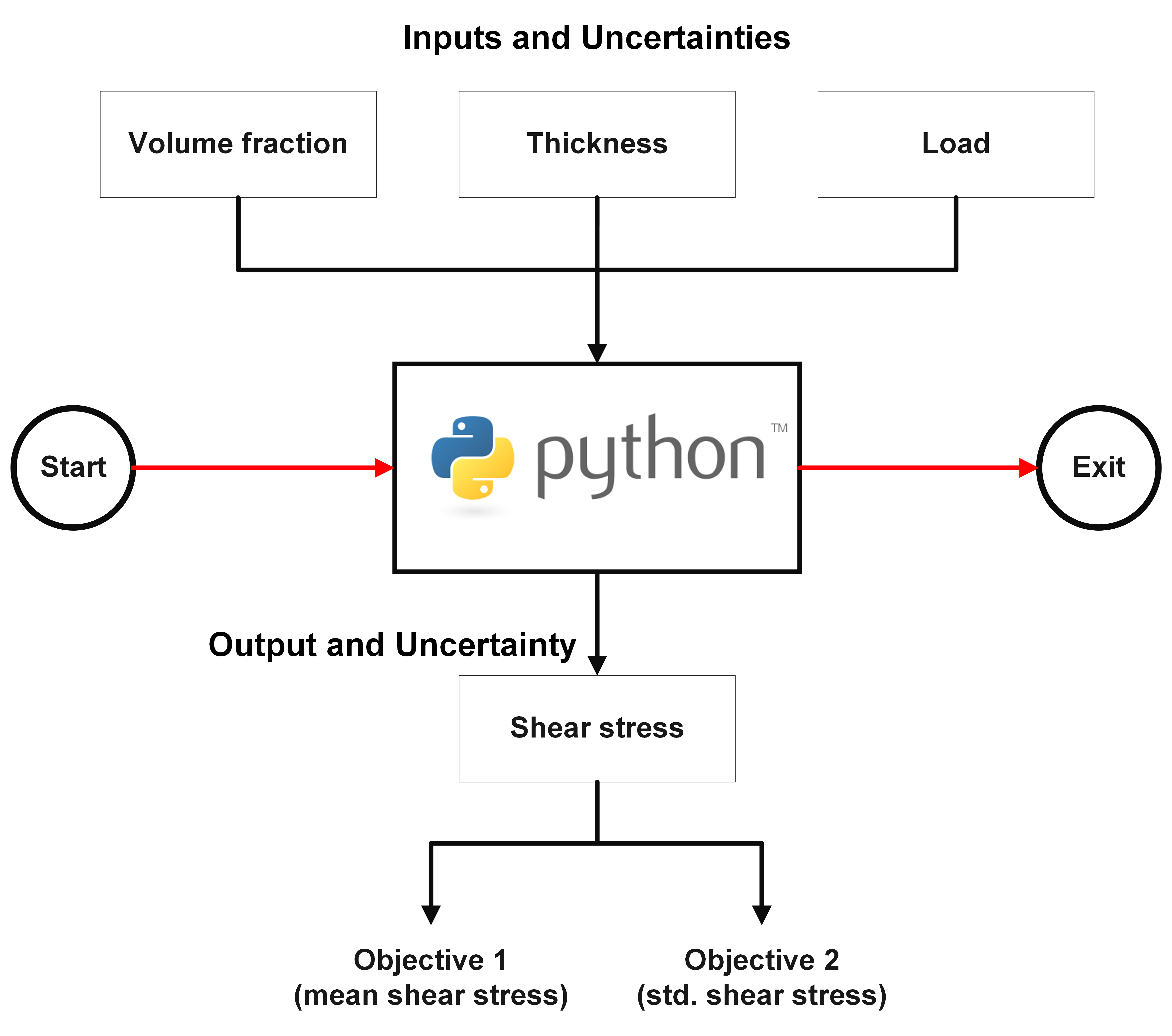}
    \caption{Workflow for robust  optimization under uncertainties. The mean and standard deviation for the output variable is set as objectives.}
    \label{fig:workflow}
\end{figure}

The data-driven methodology is shown in \ref{fig:dd} and described in the previous work \cite{kumarLUXReport}.  The workflow for the optimization is shown in Fig. \ref{fig:workflow}. The repetitive calculations for each input are performed by calling ABAQUS via a Python script. The MOGA-II algorithm is employed for the input and output datasets to optimize this problem. It is a proprietary variation of the multi-objective genetic algorithm that uses an intelligent and efficient multi-search elitism that is capable of preserving excellent solutions without unnecessarily converging to a local optimum too early on \cite{peixoto2021cfd, poles2003moga, aittokoski2008efficient, poloni1997ga, deb2000fast}. In this study, two objectives are set for the output: (1) maximize the mean value of shear stress and (2) minimize the standard deviation of shear stress.

\begin{table}[!htbp]
\caption{List of stochastic input parameters and scale classification}
\label{tb:input}
\centering
\begin{tabular}{@{}cclcccc@{}}
\toprule
Input variable  & Scale &  & Mean      & Std. (\%)          & Lower bound & Upper bound \\ \midrule
Volume fraction & Micro &  & 0.5       & \multirow{6}{*}{5} & 0.1         & 0.9         \\
Thickness 1     & Meso  &  & 1.25 (mm) &                    & 1.20        & 1.30        \\
Thickness 2     & Meso  &  & 1.25 (mm) &                    & 1.20        & 1.30        \\
Thickness 3     & Meso  &  & 1.25 (mm) &                    & 1.20        & 1.30        \\
Thickness 4     & Meso  &  & 1.25 (mm) &                    & 1.20        & 1.30        \\
Load            & Macro &  & 100 (N/mm)   &                    & 90          & 110         \\ \bottomrule
\end{tabular}
\end{table}

\begin{figure}[!htbp]
    \centering
    \includegraphics[scale=0.7]{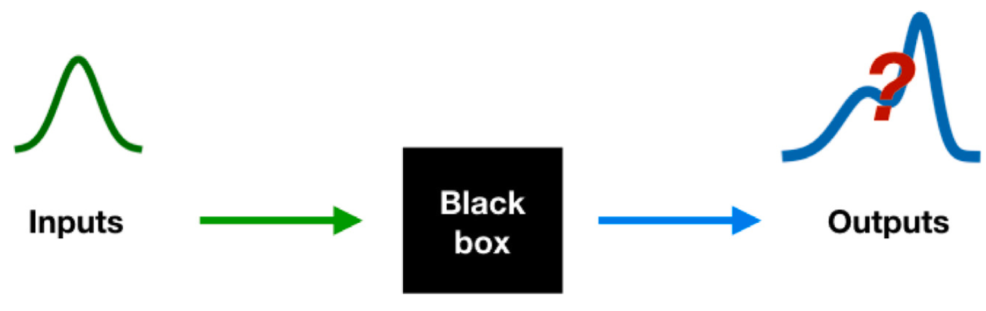}
    \caption{Propagation of uncertainties and deterministic solver as a black-box \cite{kumar2022multi}.}
    \label{fig:BBUQ}
\end{figure}

We have adopted the uncertainty propagation methodology from our previous study by Kumar \cite{kumar2022multi}. This methodology assumes that uncertainties present in the multi-scale systems can be propagated to the output response via black-box, as shown in Fig. \ref{fig:BBUQ} \cite{kumar2022multi}.

\begin{figure}[!htbp]
    \centering
    \includegraphics[scale=0.5]{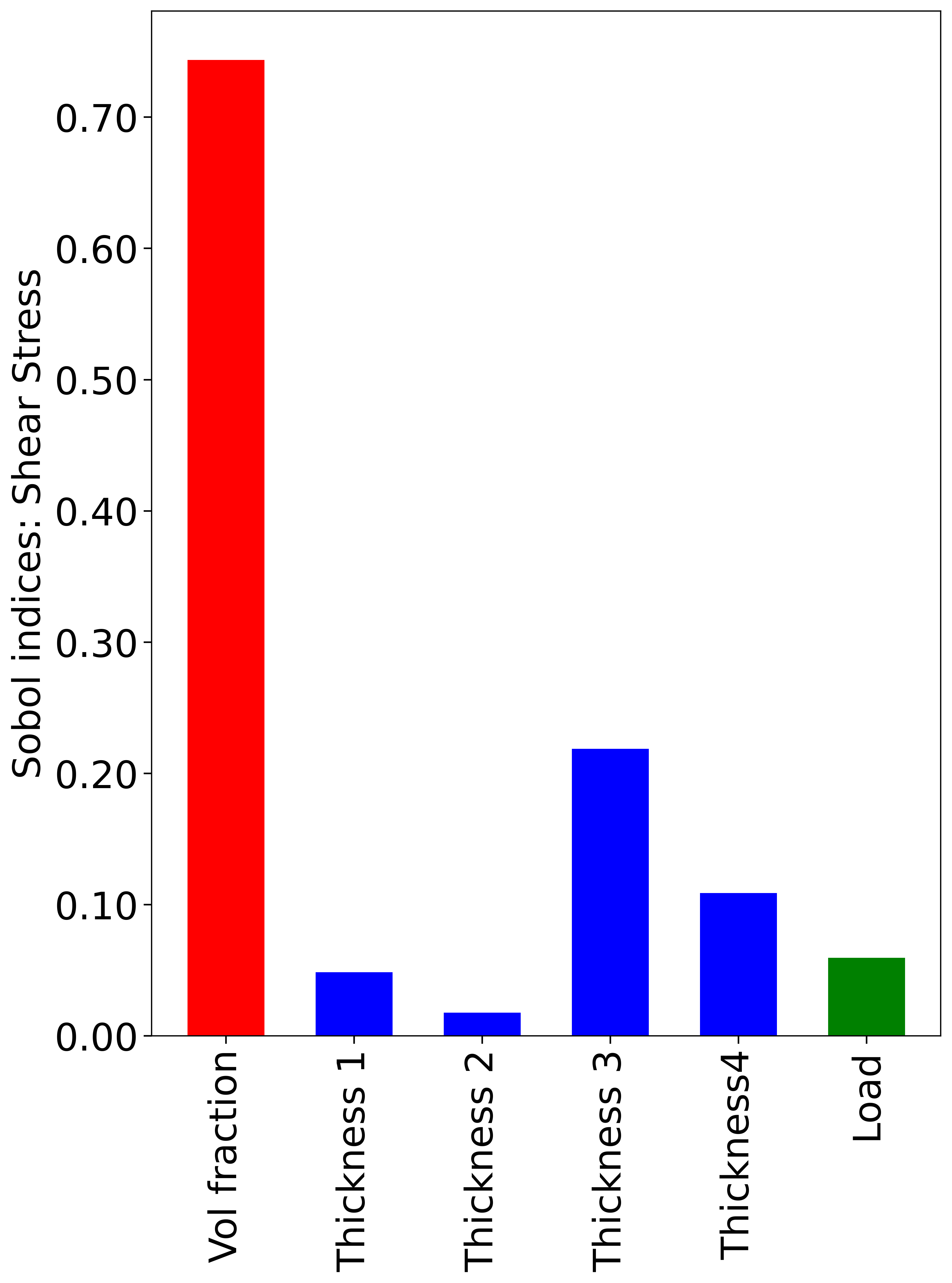}
    \caption{Result of sensitivity analysis. The y-axis represents the Sobol indices. The values of the x-axis are input variables. The colors of the plots correspond to the scale of input variables: micro (red), meso (blue), and macro (green), respectively.}
    \label{fig:sa}
\end{figure}

In order to identify the contribution of the individual input variables to the shear stress, Polynomial Chaos Expansion (PCE) based sensitivity analysis \cite{kumar2020uncertainty, Kumar2021} is performed. In Fig. \ref{fig:sa}, the y-axis represents the Sobol indices and the x-axis are input variables. It is revealed that the most affecting parameter for the shear stress is the volume fraction (micro-scale). It indicates that the contribution of variables other than micro-scale is relatively tiny.

\begin{figure}[htbp]
    \centering
    \includegraphics[scale=1]{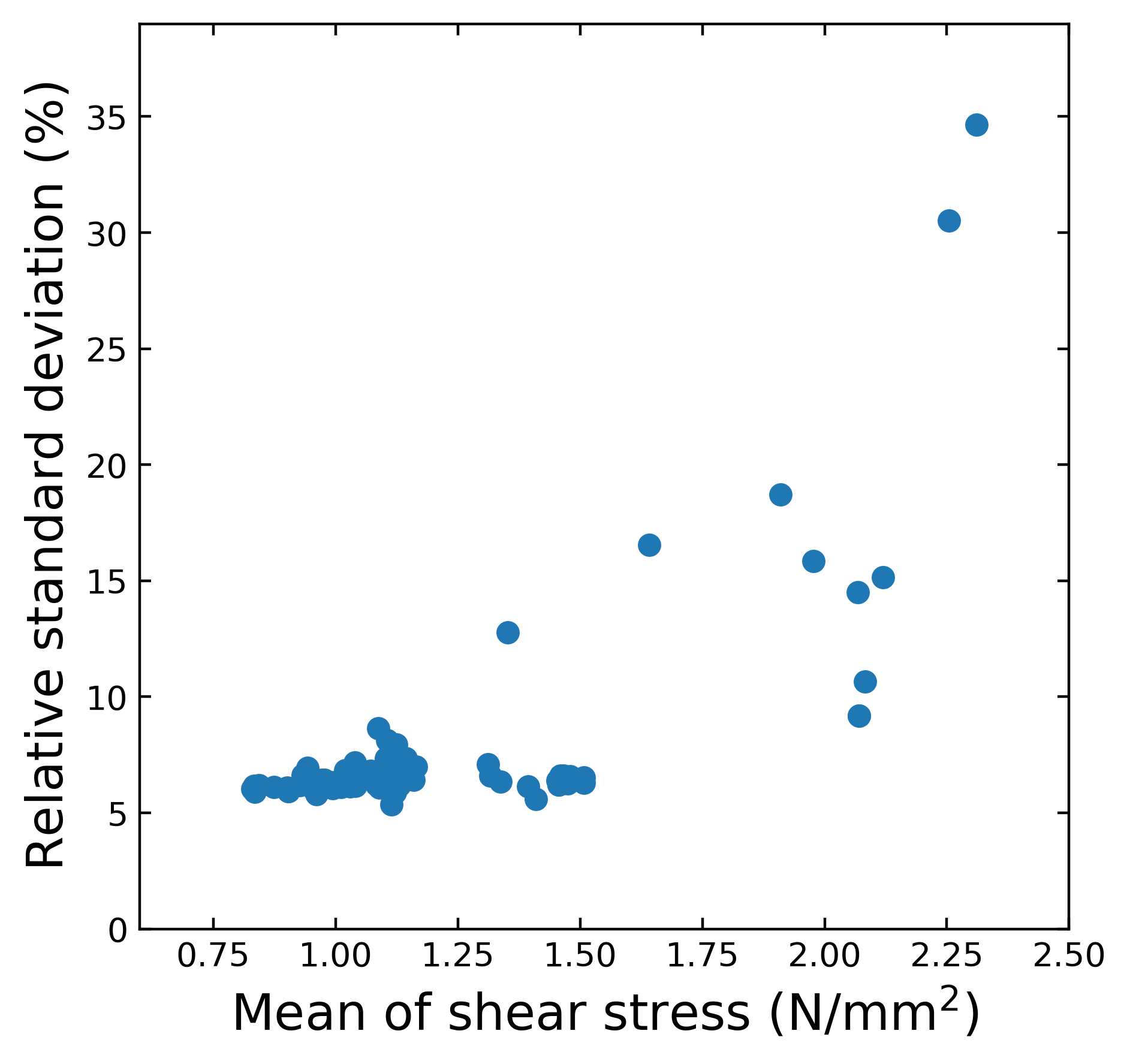}
    \caption{Correlation between the mean values of shear stress and its standard deviation.}
    \label{fig:robust_design}
\end{figure}

\begin{figure}[htbp]
    \centering
    \includegraphics[scale=0.7]{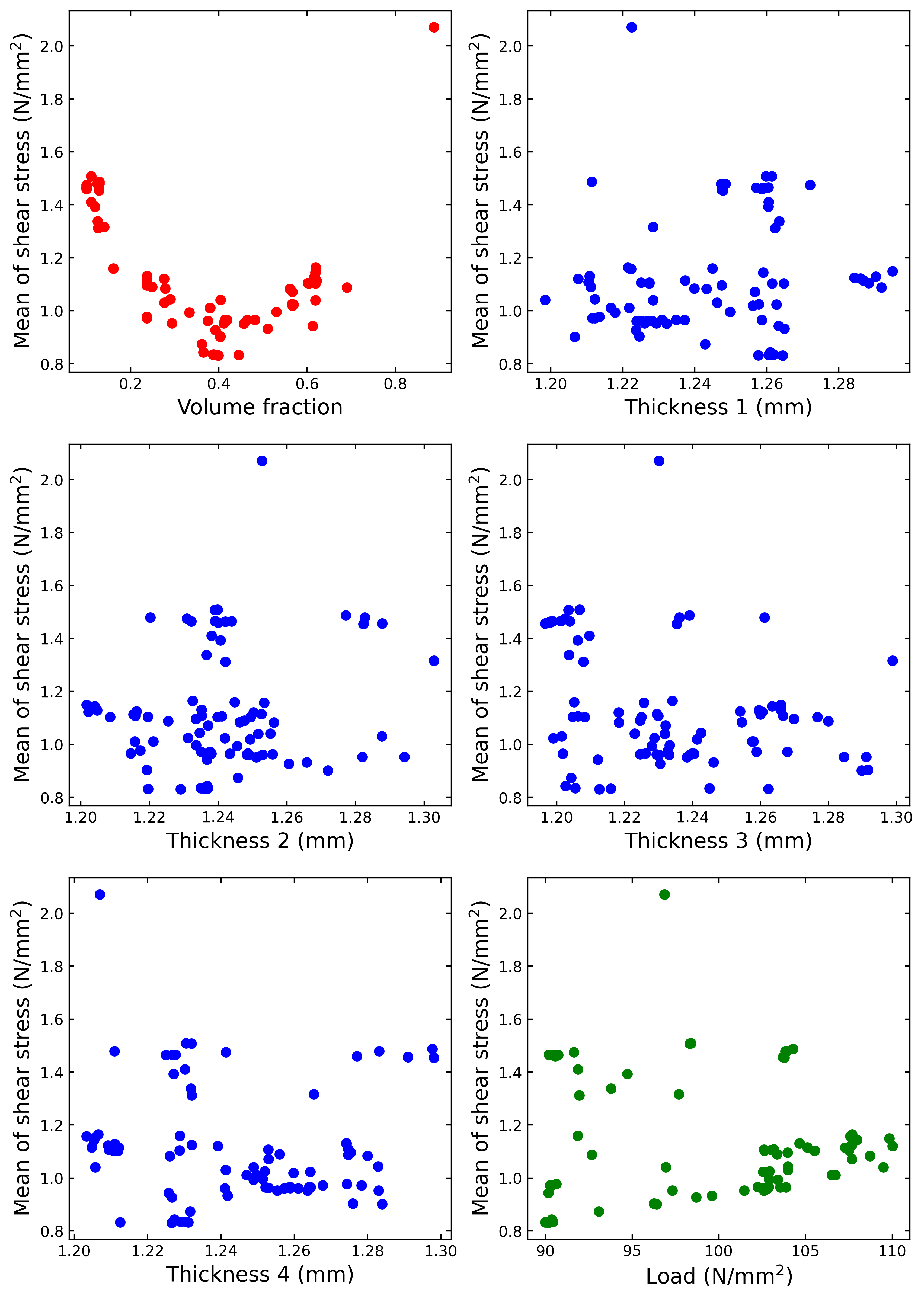}
    \caption{Correlation between the mean values of shear stress and robust design input variables. A relative standard deviation of 10\% or less for the mean of shear stress was imposed as a constraint. The colors of the plots correspond to the scale of input variables: micro (red), meso (blue), and macro (green), respectively.}
    \label{fig:robust_design_params}
\end{figure}

Robust optimization is performed. The correlation between the computed mean of the shear stress and its relative standard deviation (RSD) is shown in Fig. \ref{fig:robust_design}. In this problem, we consider increasing the output variable under the constraint RSD of less than 10\%. Fig. \ref{fig:robust_design_params} shows the correlation between the mean values of shear stress and robust design input variables under the constraint. From these results, it is found that the maximum shear stress is 2.07 $\rm N/mm^2$. The corresponding input values are the followings: volume fraction = 0.887444, thickness 1 = 1.222475 $\rm (mm)$, thickness 2 = 1.252724 $\rm (mm)$, thickness 3 = 1.230146 $\rm (mm)$, thickness 4 = 1.206935 $\rm (mm)$, and load = 96.861233 $\rm (N/mm^2)$. 

\section{Conclusion}
In product design, optimizing functionality and cost by considering uncertainties in design parameters is critical. Also, the same applies to selecting the materials that make up the product. Robust optimization and multi-scale modeling methods are two ways to address these issues. 

This chapter introduced a modeling method that combines the two techniques, using a carbon fiber structure design as a test case. This study found that the shear stress takes the maximum value of 2.07 $\rm N/mm^2$ when the input values are the followings: volume fraction = 0.887444, thickness 1 = 1.222475 $\rm (mm)$, thickness 2 = 1.252724 $\rm (mm)$, thickness 3 = 1.230146 $\rm (mm)$, thickness 4 = 1.206935 $\rm (mm)$, and load = 96.861233 $\rm (N/mm^2)$.  

The method is expected to be used in a wide range of fields, but in particular, it is expected to significantly contribute to energy systems employing innovative materials. In the major power generation methods of thermal, hydroelectric, and nuclear power, the components are exposed to a harsh utilization environment over a long period. Therefore, it is necessary to consider variable uncertainties from the design stage when optimizing its functions that vary according to their environment. In conclusion, these requirements suggest that robust optimization and multi-scale modeling will continue to develop as valuable methods.

\section*{Acknowledgement}
The computational part of this work was supported in part by the National Science Foundation (NSF) under Grant No. OAC-1919789.

\bibliographystyle{unsrtnat}
\bibliography{references}  






\end{document}